\documentclass[twocolumn,showpacs,amsmath,amssymb]{revtex4}

\usepackage[dvips]{graphicx}
\usepackage[dvips]{color}

\begin{document}
\preprint{}

\title{Evolution models with base
substitutions, insertions, deletions and selection}

\author{D.\ B.\ Saakian $^{1,2,3}$}
\email{saakian@yerphi.am}\affiliation{$^1$Institute of Physics,
Academia Sinica, Nankang, Taipei 11529, Taiwan}
\affiliation{$^2$Yerevan Physics Institute,Alikhanyan brothers St.
2,Yerevan 36,Armenia,}
 \affiliation{$^3$National Center for
Theoretical Sciences: Physics Division, National Taiwan University,
Taipei 10617, Taiwan}

\date{\today}

\begin{abstract}
The evolution model with parallel mutation-selection scheme  is
solved for the case when selection is accompanied by base
substitutions, insertions, and deletions. The fitness is assumed to
be either a single-peak function (i.e., having one finite
discontinuity) or a smooth function of the Hamming distance from the
reference sequence.
 The mean fitness is
exactly calculated in large-genome limit. In the case of insertions
and deletions the evolution characteristics depend on the choice of
reference sequence.
\end{abstract}

\pacs{87.23.Kg, 87.15.Aa}
\maketitle

\section{Introduction}

The existence of insertions and deletions is well established
experimentally. There has been a recent considerable interest in
molecular evolution models of Eigen, i.e., in connected
mutation-selection scheme [1-2], and in the parallel, i.e.,
``decoupled" mutation schemes [3-7]. The studies included mean
fitness for different fitness landscapes [4-7], and population
distributions under mutation-selection balance constraint [8]. Since
the experimental confirmation of insertion and deletion processes
[9] there have been several studies of molecular evolution models
that incorporate base substitutions, insertions, and deletions
[10-12]. In this article we integrate a concept of indels with
parallel mutation-selection processes to solve our model with
general fitness landscape and derive exact formula for the mean
fitness.

In biology research the term indel stands for either
insertion alone or deletion alone or both these processes
present simultaneously. Indels play an important role in
phylogenic analysis in practical population
genetics [13], where incorrect handling of indels may give
unrealistic outcomes.

In the parallel mutation-selection model any genotype
configuration $i$ is specified as a sequence of $N$
two-valued letters (alleles) $s_n=\pm 1$, $1 \le n \le N$.
We denote such configuration $i$ by
$S_i\equiv ({s_1^i, \dots, s_N^i})$. Probability $p_i$ that
configuration $S_i$ occurs in genome, $1 \le i \le 2^N$,
satisfies
%%%%% equation 1
\begin{equation}
\label{e1}
\frac{{dp}_i}{dt}={p_i} \left( r_i- \sum_{j=1}^{2^N}r_j
p_j \right) + \sum_{j=1}^{2^N}\mu_{i j}p_j  ,
\end{equation}
%%%%%
where $r_i$ is the fitness, and $\mu_{ij}$ is the mutation rate from
$S_i$ to $S_j$ per unit time. For the Crow-Kimura model [4]:
$\mu_{ij}= -a N$ if the Hamming distance $d_{ij}$ is zero,
$\mu_{ij}= a$ if $d_{ij}=1$, and $\mu_{ij}= 0$ if $d_{ij}>1$; where
$d_{ij}=(N-\sum_n s_n^i s_n^j)/2$.

In the models studied here we consider the following
three independent parallel processes
in the genome: base substitutions, deletions and insertions.
Assuming constant genome-variation rates per site, we denote
$a/N_0$, $b/N_0$, and $c/N_0$ the rates of mutation, insertion,
and deletion, respectively, where $N_0>>1$ is the scale length of
the genome. Unlike in the well-studied cases of the parallel
mutation-selection scheme and the Eigen model, now the genome
length can be varied. In this paper we focus only on symmetric
fitness landscape, i.e., when the fitness of the genome is a
function of Hamming distance from one reference genome sequence.
The fitness is assumed to be either a single-peak function (i.e.,
having one finite discontinuity) or a smooth function of the
Hamming distance.

In the first model, analyzed in Sec.\ref{ordered}, we are
investigating indels acting in a toy problem when the reference
sequence is ordered, i.e., when it contains only one letter (either
$+1$ or $-1$) at all positions. Obtaining solution to this toy
problem is by no means trivial because neither the maximum principle
[5] nor the Hamilton-Jacobi method [8] can be applied directly. A
more realistic case is analyzed in Sec.\ref{random} for random
reference sequence when the letters $+1$ and $-1$ are randomly
distributed along genome length. For symmetric fitness in parallel
mutation-selection model without indels the choice of the reference
sequence does not affect the solution, which is a consequence of the
existing symmetry of the governing equations. The introduction of
indels to the model breaks this symmetry and the effect of indels
acting on sequence space is to change the solution. This change
depends on the choice of the reference sequence. If we choose as the
reference sequence the one with all $+$
 alleles, the result of deletion is the same for all the N possible
 position of deleted allele.
 In case a random reference sequence we have different results (sequences)
 after different positions of deleted allele.
In our model, an individual indel event means either insertion or
deletion of a single letter in the genome sequence, one at a time,
but there may be many indel events during evolution. In this article
we focus on investigating a ``successful selection" phase, i.e., the
phase (range of parameters) with majority of population being
localized around the reference sequence. Our results are discussed
in Sec.\ref{discussion}.

\section{Ordered reference sequence \label{ordered}}

We choose the reference sequence that has all the alleles $+1$ and
the initial distribution of sequences that is symmetric under
permutations. Individual configuration is denoted by $(N,L)$, where $N$
is genome length, and $L$ is the number of ($+1$)-alleles
in the configuration. The fitness is $N_0r(N,L)$.
Considering only one-letter deletion or insertion at a time
there may be three processes that start at $(N',L')$ and
end at $(N,L)$:
\begin{itemize}
\item Simple base substitutions at the rate of $a/N_0$.
At the beginning there are $L$ configurations
$(N,L-1)$ and $N-L$ configurations with $(N,L+1)$.
\item Deletions at the rate of $b/N_0$.
At the beginning there are $N+1$ configurations
$(N+1,L+1)$ and $N+1$ configurations with $(N+1,L)$.
\item Insertion at the rate of $c/N_0$.
At the beginning there are $L$ configurations
$(N-1,L-1)$ and $(N-L)$ configurations $(N-1,L)$.
\end{itemize}
During base substitutions the letters (alleles)
change their signs. During deletion one of the letters
disappears. During insertion a new letter (either $+1$ or $-1$)
is added randomly at any of the $(N+1)$ positions along
the chain.

There are two interests in solving this model, usually treated
separately. One interest concerns to genome growth [10-12]. The
other interest is the study of successive selection phase, which we
present here for the case when selections and base-substitutions are
accompanied by deletions and insertions.

The occurrence probability $p(N,L)$ denotes a fractional
number of configurations $(N,L)$ in the population.
For symmetric fitness landscape and permutation-symmetric
initial distribution, probabilities $p(N,L)$ satisfy
the equations \cite{ss82}:
%%%%% equation 2
\begin{equation}
\label{e2}
\begin{aligned}
&\frac{dp(N,L)}{dt}= \\
& p(N,L) \left( N_0 \, r(N,L)-
\frac{N}{N_0}(a+b)- c \, \frac{N+1}{N_0} \right) \\
&+a \left( \frac{L}{N_0}p(N,L-1)+p(N,L+1)\frac{N-L}{N_0} \right) \\
&+b \left[ p(N+1,L+1)+p(N+1,L) \right] \frac{N+1}{N_0} \\
&+\frac{c}{2} \left( p(N-1,L-1)\frac{L}{N_0}+p(N-1,L)\frac{N-L}{N_0} \right) \\
&-p(N,L)\sum_{N',L'}r(N',L')p(N',L') \binom{N'}{L'}.
\end{aligned}
\end{equation}
%%%%%
In case of symmetric fitness landscape,  for finding steady-state
mean fitness it is sufficient to consider only symmetric evolution.
We solved Eq.(\ref{e2}) numerically, varying genome length between
$N_1$ and $N_2$ subject to $N_2-N_1\gg 1$. The numerical results for
two values of genome length are presented in Fig.~1.

%%%%% figure 1
\begin{figure}
\large \unitlength=0.1cm
\begin{picture}(80,110)
\put(-26,56){\includegraphics{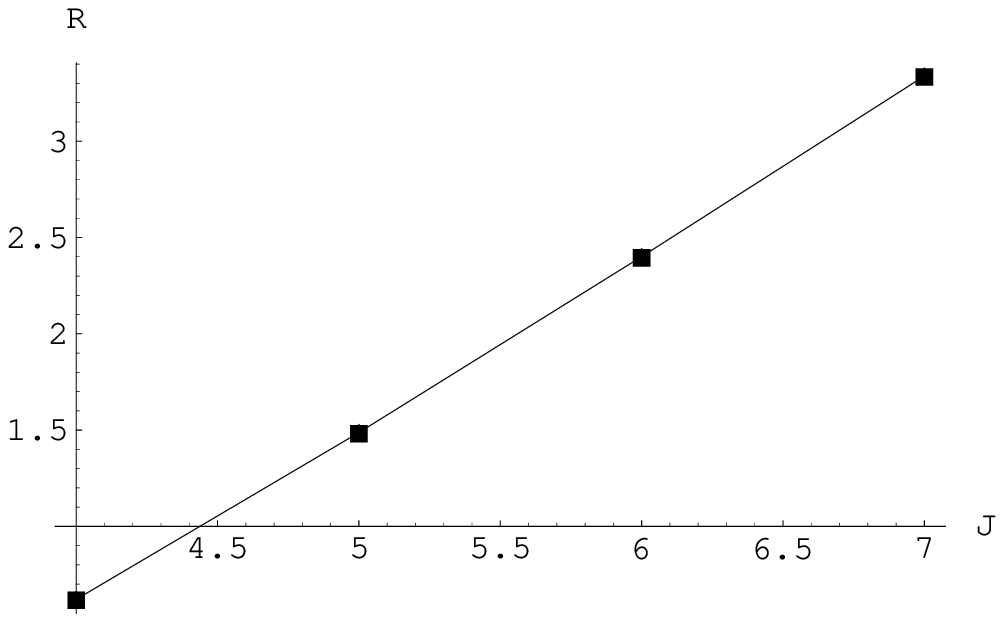}}
\put(-26,0){\includegraphics{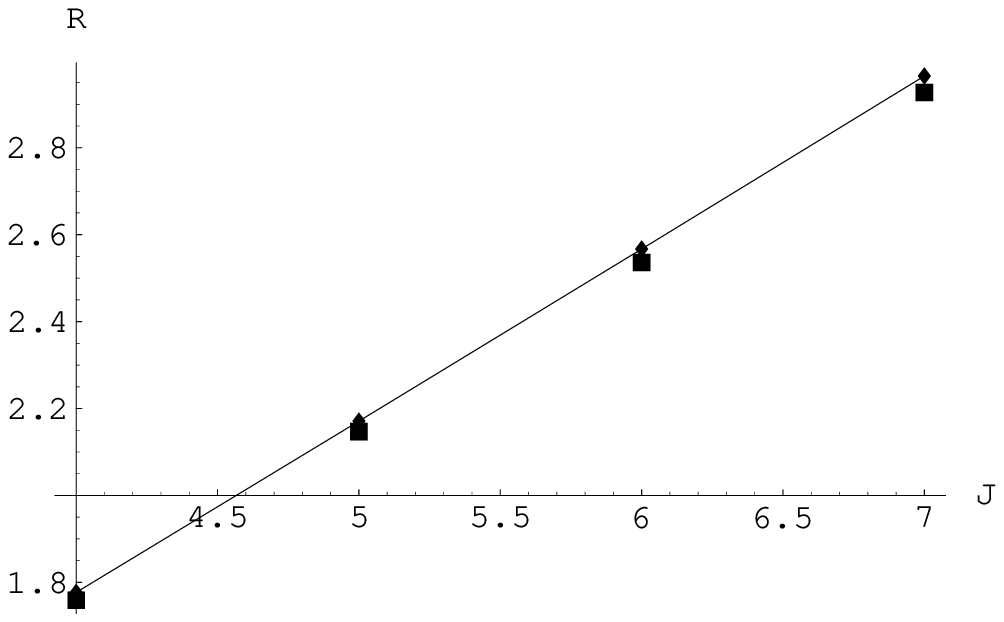}}
\put(20,100){\small{(a)}} \put(20,44){\small{(b)}}
\end{picture}
\caption{
Mean fitness $R$ vs fitness parameter $J$ for
$a=b=1$, and $c=2$. Theoretical
results are plotted as continuous lines. Numerical results
are represented by symbols. Error bars give
percent difference between numerical and theoretical results.
(a) Single-peak mean fitness for $N=1000$.
Error bars are about $0.05 \%$, smaller than symbol size.
(b) Quadratic fitness, $r=Jm^2$, for $N=200$.
Error bars are about $0.5 \%$.
}
\end{figure}
%%%%%
Eq. (2) is slightly modified near the border values of N, as at
$N_2$ there are only deletions and at $N_1$-only insertions.
 The
weighted sum over all equations is zero, where the weights $(^N_L)$
are numbers of configurations with the same $N$ and $L$.

 For
single-peak fitness landscape we set all fitness values to zero but
one:
%%%%% equation 3
\begin{equation}
\label{e3} r(N_0,N_0)=J, \quad r=0 \; \; \mathrm{otherwise},
\end{equation}
and $N_0=\frac{N_1+N_2}{2}$.
%%%%%
In continuous-time model, considered in this work, fitness landscape
Eq.(\ref{e3}) can be rescaled by an additive constant, which is a
standard procedure in statistical physics. In discrete-time models
all fitness values would have to be positive. For base substitutions
acting alone without indels, with the choice given by Eq.(\ref{e3})
the system of Eqs.(\ref{e2}) decouples, which leads to a single
equation for only one master-type  (reference sequence) probability
[6] from which other probabilities are obtained recursively. When
base substitutions and indels are simultaneously acting, for
single-peak fitness defined by Eq.(\ref{e3}) the system
Eqs.(\ref{e2}) does not decouple but, nonetheless, can be reduced to
a tractable problem that can be treated analytically. The reduction
procedure is outlined in the next paragraphs.

We consider the following scaling of Eqs.(\ref{e2}):
%%%%% equation 4
\begin{equation}
\label{e4}
\begin{aligned}
p(N,N) &\sim 1,\\
p(N,L) &\sim 1/N_0^{N-L}.
\end{aligned}
\end{equation}
%%%%%
Since the scaling (\ref{e4}) suppresses contributions from all terms
$(N,L)$ for $L<N$ as $1/N$, after the scaling we obtain a complete
set of equations for the class $p(N,N), N_1\le N\le N_2$ of
frequencies. For example, for $p(N,N-1)$ we derive from
Eqs.(\ref{e2}):
%%%%% equation 5
\begin{equation}
\label{e5}
p(N,N-1)=\frac{a}{N_0}p(N,N)+\frac{c}{2N_0}p(N-1,N-1),
\end{equation}
which is easily verified to be consistent with the scaling
ansatz (\ref{e4}).

Denoting by $\vec P$ a collection of all $p(N,N)$, where
$P_l=p(N_1-1+l,N_1-1+l)$, we write Eq.(\ref{e2}) for $p(N,N)$ as
$$\frac{d\vec P}{dt}=\hat A \vec P-R\vec P,$$
where $R$ is the mean fitness, $R=Jp(N_0,N_0)$; and,
the elements of matrix $\hat A$ are
%%%%% equation 6
\begin{equation}
\label{e6}
\begin{aligned}
A_{ll} &= -(a+b+c)+J \, \delta_{l,l_0}, \\
A_{l,l+1} &= b,\\
A_{l,l-1} &= c/2,
\end{aligned}
\end{equation}
%%%%%
where $l_0=N_0-N_1+1$, and $\delta_{l,l_0}$ is Kronecker's symbol.
In successful-selection phase the majority of population is
distributed around the configuration $(N_0,N_0)$.
The details of matrix $\hat A$ near the borders at $l=1$,
$l=M+1$, and $M\equiv N_2-N_1$, are irrelevant for the computation
of the mean fitness in the successful-selection phase in the sense that
the mean fitness is insensitive to variations in these border
values. The mean fitness $R$ is obtained in the standard way as
the largest eigenvalue of $\hat A$ by solving the secular equation
%%%%% equation 7
\begin{equation}
\label{e7} \det \left( \hat A- \lambda \hat I \right) =0,
\end{equation}
%%%%%
where $\hat I$ is the identity matrix, and $R= \max{(\lambda)}$.

To calculate $R$ within the $1/N_0$-accuracy we utilize the
properties of determinant and in Eq.(\ref{e7}), modify the matrix
$A$, taking $A_{1,M+1}=b$ and $A_{M+1,1}=c/2$. Then, we define an
auxiliary function $g(J)$ by $g(J) \equiv \det (\hat A - R \hat I
)$. Since $g(J)$ is linear we write
$$g(J)=g(0)+ J g'(0),$$
where $g(0) = \det ( \hat B - R \hat I)$; the matrix $\hat B$
is the value of the matrix $\hat A$ computed at $J=0$;
and, $g'(0)$ is the first derivative of $g(J)$ computed
at $J=0$. Because matrix $\hat B$ is symmetric and cyclic it is
relatively straightforward to write $g(0)$ and $g'(0)$ explicitly:
%%%%% equation 8
\begin{equation}
\label{e8}
\begin{aligned}
&g(0)=
 \prod_{l=0}^{M} \left( b e^{i 2 \pi l/M }+
\frac{c}{2} e^{-i 2 \pi l/M} -(a+b+c)-R \right) ,\\
&\frac{g'(0)}{g(0)}=
\frac{1}{M}\sum_{l=0}^{M} \frac{1}
{b e^{i 2 \pi l/M}+\frac{c}{2} e^{-i 2 \pi l/M}-(a+b+c)-R}.
\end{aligned}
\end{equation}
%%%%%
In the thermodynamic limit of large $N_0$, $N_1$, and $N_2$,
the infinite summation on the right-hand-side of
Eq.(\ref{e8}) becomes a contour integral in complex plane.
The left-hand-side of Eq.(\ref{e8}) is
$g'(0)/g(0) = (g(J)/g(0) - 1)/J$ and $g(J)=0$ because
of Eq.(\ref{e7}). Thus, making the substitution $z= \exp(i2 \pi l/M)$,
Eq.(\ref{e8}) gives the relation between the mean fitness $R$ and
the fitness $J$ of the peak configuration:
%%%%% equation 9
\begin{equation}
\label{e9}
\begin{aligned}
1 &= -\frac{J}{2\pi i}\oint \frac{dz}{z}\frac{1}{b
z+\frac{c}{2z}-(a+b+c)-R}\\
&= \frac{J}{\sqrt{(R+a+b+c)^2-2bc}}.
\end{aligned}
\end{equation}
%%%%%
Inverting Eq.(\ref{e9}) gives the mean fitness $R$ and
fractional population $P_m$ of the peak configuration:
%%%%% equation 10
\begin{eqnarray}
\label{e10}
R=\sqrt{J^2+2bc}-(a+b+c), \\
P_m\equiv p(N_0,N_0)=\frac{\sqrt{J^2+2bc}-(a+b+c)}{J} , \nonumber
\end{eqnarray}
%%%%%
and the error-threshold condition
%%%%% equation 11
\begin{equation}
\label{e11}
J\ge\sqrt{(a+b+c)^2-2bc}.
\end{equation}
%%%%%
Results of Eqs.(\ref{e3}) and (\ref{e10}) are
illustrated in Fig.~1a.

\subsection{Nonzero fitness at one $N$ value\\and many $L$
values \label{orderedA}}

The sharp-peak fitness defined by Eq.(\ref{e3}) is
an oversimplification as it is believed that realistic
fitness landscapes are highly complicated and irregular.
As a step towards generalization we consider now
fitness that is nonzero at only one $N$ value, set to $N=N_0$,
and at many values of $L$. Here, $L$ is the number of
the ($+1$)-alleles in the genome and Hamming distance to
the reference configuration is $N-L$. For this more general
fitness we take
%%%%% equation 12
\begin{equation}
\label{e12}
r(N,L)=\delta_{N,N_0} \, f(2L/N_0-1),
\end{equation}
%%%%%
where $f(\cdot)$ is a smooth function.
Following a method introduced by Baake and Wagner \cite{bw01}
we transform Eqs.(\ref{e1}) to a more convenient form with
the use of the substitution
%%%%% equation 13
\begin{equation}
\label{e13} y(N,L)=p(N,L)\sqrt{\frac{N!}{L!(N-L)!}}.
\end{equation}
%%%%%
Equations for the weighted fractional populations $y(N,L)$ simplify
in the large-genome limit. For the computation of the mean fitness
they are easier to handle than the original Eqs.(\ref{e1}). We have
checked rigorously by calculating the distribution $y(N,L)$ that it
is a smooth function of $L/N_0$ for the given $N=N_0$, although it
is not smooth for all $N$. Assuming that $y(N,L)$ is a smooth
function of $(2L/N-1)$ near $N=N_0$ and near the location $L_0(N_0)$
of its maximum, we replace $y(N,L)$ with $y(N,L_0(N_0))$ in the
coupled system of equations for $y(N,L)$ that was obtained from
Eqs.(\ref{e1}) after applying transformation (\ref{e13}). As
described for single-peak fitness, this gives a partial decoupling.
For the decoupled part we have the eigenvalue problem
$$\hat A \vec y = \lambda \vec y,$$
where $y_l=y(N,L_0(N_0))$, $l=N-N_1+1$, and $R= \max{(\lambda)}$.
Again, the matrix $\hat A$ is tri-diagonal. In the limit of large
$N$ the eigenproblem for $\hat A$ gives
%%%%% equation 14
\begin{equation}
\label{e14}
\begin{aligned}
\lambda \, y_l &=
y_l \left( \delta_{l,l_0} \, f(m)- (a+b+c)+a\sqrt{1-m^2} \right) \\
&+ \left( y_{l+1}b
+y_{l-1} \frac{c}{2} \right) \frac{\sqrt{1+m}+\sqrt{1-m}}{\sqrt{2}},
\end{aligned}
\end{equation}
%%%%%
where $m=(N-2L_0)/N_0,l_0=N_0-N_1+1$. There is full analogy between
this problem and the problem already solved for single-peak
function. Quadratic form for the single-peak problem can be obtained
from quadratic form for the current problem by performing the
mapping: $b \to b\frac{\sqrt{1-m}+\sqrt{1+m}}{\sqrt{2}}$, $c \to
c\frac{\sqrt{1+m}+\sqrt{1-m}}{\sqrt{2}}$, $R \to R-a\sqrt{1-m^2}$.
By repeating the steps that lead to Eq.(\ref{e9}) we derive
%%%%% equation 15
\begin{equation}
\label{e15}
\begin{split}
R= \max_m \, \{
&- (a+b+c) +a\sqrt{1-m^2} \\
&+ \sqrt{f^2(m)+ bc \, (\sqrt{1+m}+\sqrt{1-m})^2} \, \} .
\end{split}
\end{equation}
%%%%%
Theoretical results of Eqs.(\ref{e12}) and (\ref{e15}) for
quadratic fitness are presented in Fig.~1b.

\subsection{General fitness landscape \label{orderedB}}

For the ordered reference sequence we assume now
nonzero fitness at many $N$ and $L$ values, i.e.,
fitness is a function of both genome length
and the number of $(+1)$-alleles:
%%%%% equation 16
\begin{equation}
\label{e16}
r(N,L)=f(\frac{N}{N_0},\frac{2L-N}{N_0}).
\end{equation}
%%%%%
We assume that fitness $f(n,m)$ is a smooth function of both its
arguments, i.e., it is also smooth in genome length. This is in
contrast to the model of Sec.\ref{orderedA} where the fitness may
change drastically even within one unit of genome length. As in the
previous two examples, to find the mean fitness $R$ it is requested
in our approach that distribution $y(n,m)$ must have a maximum
localized at $N$ and $L(N)$. Then, if the maximum exists the system
of equations Eqs.(\ref{e1}) for fractional populations can be
partially decoupled at configuration $(N,L(N))$. This leads to the
algebra problem of finding the largest eigenvalue of a matrix,
$R=\max(\lambda)$. In analogy with Eq.(\ref{e14}), the intermediary
result is
%%%%% equation 17
\begin{equation}
\label{e17}
\begin{split}
\lambda \, y_l &=  \\
&y_l \left( \delta_{l,l_0} \, f(n,m)-n(a+b+c)+a\sqrt{n^2-m^2} \right)  \\
&+ \left( y_{l+1}+y_{l-1} \right)
\left( \sqrt{n+m}+\sqrt{n-m} \right) \frac{\sqrt{bc}}{2}.
\end{split}
\end{equation}
%%%%%
Finally, the mean fitness $R$ is the largest eigenvalue
$\lambda$ that is obtained by solving Eqs.(\ref{e17}):
%%%%% equation 18
\begin{equation}
\label{e18}
\begin{split}
R =
\max_{n>m} \, \{ &f(n,m)-n(a+b+c)+a\sqrt{n^2-m^2} \\
&+(\sqrt{n+m}+\sqrt{n-m})\sqrt{bc} \, \}.
\end{split}
\end{equation}
%%%%%
Note, the final result Eq.(\ref{e18}) requires finding the
maximum in two-dimensional space of arguments $n$ and $m$.

\section{Random reference sequence \label{random}}

In this model a reference sequence contains both
$+1$ and $-1$ alleles that are randomly distributed
along the entire genome length. Single-peak fitness
is analyzed in the next paragraph, followed by the
extension of the model to general symmetric
fitness.

When reference configuration contains a number of pieces
with consecutive $l$ number of all $(+1)$-alleles or
all $(-1)$-alleles per piece, the mean fitness scales
according to
%%%%% equation 19
\begin{equation}
\label{e19} R \sim (l/N_0)^{\alpha}, \: \mathrm{where} \quad
\alpha\ge 1.
\end{equation}
%%%%%
Since for the random reference sequence we have $l/N_0\ll 1$,
Eq.(\ref{e19}) implies that in this case the number of
configurations that return to the initial reference sequence is
negligible and can be neglected. The final result is exactly the
same as though there were only base substitutions with the rate
$a+b+c$:
%%%%% equation 20
\begin{equation}
\label{e20} R=J-(a+b+c),
\end{equation}
%%%%%
and error threshold condition is
\begin{equation}
\label{e21} J>(a+b+c)
\end{equation}

 For continuous-time Eigen's model,
where $r_1=A = \mathrm{const}$ and $r_i=1, i>1$, the presence of
indels modifies error threshold to:
%%%%% equation 21
\begin{equation}
\label{e22} QA\ge 1 ,
\end{equation}
%%%%%
where $Q$ is the probability of errorless reproduction of
the entire genome.

In general-symmetric-fitness model the fitness is defined by
Eq.(\ref{e16}), where we replace $L$ by a parameter $N-d$ that
explicitly contains Hamming distance $d$ between the current
configuration and  reference configuration. Here, for any genome
length $N$ there is its reference configuration, thus, the Hamming
distance $d$ is the second independent parameter in the definition
of general fitness:
%%%%% equation 22
\begin{equation}
\label{e23} r(N,d)=f(\frac{N}{N_0},\frac{N-2d}{N_0}),
\end{equation}
%%%%%
where $f(\cdot,\cdot)$ is assumed to be a smooth
function of two parameters. By the method outlined in
Sec.\ref{ordered} we obtain the following general result
for the mean fitness
%%%%% equation 23
\begin{equation}
\label{e24} R = \max_{n>m}\left\{ f(n,m)-n(a+b+c)+a\sqrt{n^2-m^2}
\right\} .
\end{equation}
%%%%%

\section{Discussion \label{discussion}}

The toy model with ordered reference sequence and single-peak
fitness, studied in Sec.\ref{ordered}, is an interesting case from
methodological point of view. It describes the evolution when there
is a desperate difference between genome length $N$ and the number
$L$ of $(+1)$-alleles in fractional-population distribution.
Genotype frequencies for this class are smooth functions of $L$ in
the neighborhood of the peak distribution. Error-threshold
condition, Eq.(\ref{e11}), depends on all the rates in a nonlinear
fashion because of the existence of reversal processes introduced by
indels, which processes may drive the evolution towards the
reference sequence. This picture is unlike to what we learn from
Crow-Kimura (parallel) model without indels, where error threshold
depends linearly on the rates. In the generalized model with
single-peak fitness, studied in Sec.\ref{random}, where the
reference sequence is a random sequence of $+1$ and $-1$ alleles,
the error-threshold formula, Eq.(\ref{e21}), simplifies again to
that for the Crow-Kimura model with an efficient base substation
rate.

General symmetric-fitness model with random reference sequence,
analyzed in Sec.\ref{random}, presents most realistic situation
where both genome length and reference sequence are allowed to vary.
In this work we investigated only steady-state characteristics of
this model.

In our computational approach we used standard methods
of linear algebra to partially decouple a system of
evolutionary equations around the peak distribution
and find the mean fitness as the largest eigenvalue
of the decoupled subsystem. As the excellent agreement
between our analytical and numerical solutions demonstrates
(see Fig.~1) our methodology has a promise of becoming
a routine approach in solving general evolutionary
problems with varying genome length, alongside the methods
of quantum mechanics \cite{bbw97} that are good when genome
length is fixed, and the methods of quantum field theory
\cite{sa03,ge07}.

Haploid models with indels that were studied in this work can be
extended to diploid evolution models with parallel insertions and
deletions. Similar complex models were already considered to study
the evolution of gene families via conversion processes \cite{na84}
and gene crossover processes \cite{oh80}. The latter mentioned
mechanisms could be in principle handled analytically by modern
methods \cite{sa08} that proved to be successful in treating diploid
evolution \cite{ba92,sp99}.

In evolution research the models often ignore either
selection processes [10-12] or indels [2-7], however,
it is generally accepted that the concurrent selection
and indels play an important role in biology. The models
that are capable to describe these two processes as
acting simultaneously could give connection with the
phylogeny analysis and with the investigation of gene
families. Our introductory study of this work shows
that it is possible to analytically derive some class
of results when selection is accompanied by indels.

In summary, in this work we introduced a method that
allows one to investigate a broad class of evolution
models. We solved parallel mutation-selection model
with general symmetric-fitness landscapes in case of
simultaneously acting base substitutions, insertions
and deletions. Our findings indicate that in the steady
state of this evolution model the mean fitness depends
strongly on the choice of the reference sequence.

\acknowledgments

I thank J.~Felenstein, A.~Kolakowska, and H.C.~Lee for
useful discussions. This work has been supported by
the Volkswagenstiftung grant ``Quantum Thermodynamics,''
by the National Center for Theoretical Sciences in Taiwan,
and by Academia Sinica (Taiwan) under Grant No. AS-95-TP-A07.

\end{document}